\documentclass[prl,twocolumn,showpacs]{revtex4}

\usepackage{graphicx}
\usepackage{dcolumn}
\usepackage{bm}
\usepackage{amsfonts}
\usepackage{epsfig}
\usepackage{multirow}

\newcommand{\R}{\mathbb R}

\begin{document}

\title{Discrepancy between sub-critical and fast rupture roughness: a cumulant analysis}

\author{N. Mallick, P.-P. Cortet, S. Santucci}
\thanks{Current address : Fysisk Institutt, Universitetet i Oslo, PostBoks 1048 Blindern, 0316 Oslo,
Norway} \author{S.G. Roux, L. Vanel}\email{Loic.Vanel@ens-lyon.fr}

\affiliation{Laboratoire de physique, CNRS UMR 5672, Ecole Normale
Sup\'erieure de Lyon, 46 all\'ee d'Italie, 69364 Lyon Cedex 07,
France}

\pacs{68.35.Ct, 62.20.Mk, 02.50.-r}


\begin{abstract}
We study the roughness of a crack interface in a sheet of paper. We
distinguish between slow (sub-critical) and fast crack growth
regimes. We show that the fracture roughness is different in the two
regimes using a new method based on a multifractal formalism
recently developed in the turbulence literature \cite{Delour}.
Deviations from monofractality also appear to be different in both
regimes.
\end{abstract}

\maketitle


%
%

Since the early description of rough fractures as self-affine
surfaces \cite{MandelbrotBarabasi}, the existence of universal
roughness exponents has been strongly debated
\cite{Bouchaud9097Maloy92}. There are now many experimental
evidences for a non-unique value of the roughness exponent of
fracture surfaces. Different exponents can be found due to the
anisotropy of the fracturation process \cite{BouchbinderPonson}, the
anisotropy of the material structure \cite{Menezes} or anomalous
scaling related to finite-size effects \cite{Lopez}. A recent
observation also shows that, in rupture of paper, the crack
interface can not be described with a single roughness exponent and
would be multifractal \cite{Bouchbinder06}.

Most of the time, roughness exponents appear independent of crack
velocity. For rather slow velocities, no effect of the velocity on
roughness has been observed in plexiglas
($v=10^{-7}-5\,10^{-5}\,\rm{m.s}^{-1}$), glass
($v=10^{-9}-5\,10^{-8}\,\rm{m.s}^{-1}$), intermetallic alloys
($v=10^{-8}-5\,10^{-5}\,\rm{m.s}^{-1}$) or sandstone
($v=10^{-4}-10^{-2}\,\rm{m.s}^{-1}$)
\cite{Schmittbuhl97Daguier97Boffa99}. On the contrary, in dynamic
fracture of plexiglas ($v \gtrsim 600\,\rm{m.s}^{-1} \simeq
0.45\times$ Rayleigh wave speed), variations of the roughness
exponents with the velocity have been observed \cite{Boudet}. As
pointed out in a recent review \cite{Alava}, there is a lack of
experimental studies concerning the influence of fracture kinetics
on roughening.

In this Letter, we study the roughness of a crack interface in a
sheet of paper \cite{HorvathSalminen} for which multifractality was
observed \cite{Bouchbinder06}. We show that it is crucial to take
into account the non-stationarity of crack growth. Schematically,
crack growth occurs in two different regimes: a sub-critical regime
where the growth is slow ($v=10^{-5}-10^{-2}\,\rm{m.s}^{-1}$) and a
regime where the crack growth is fast ($v\sim 300\,\rm{m.s}^{-1}$).
We show that the fracture roughness is different in the two regimes.
Various method of analysis of fracture roughness have been proposed
\cite{Schmittbuhl95}. Here, we introduce a new method based on a
multifractal formalism recently developed in the turbulence
literature \cite{Delour}. This formalism allows us to measure
reliably scaling exponents in fracture and to quantify very
precisely deviations from monofractal behavior.

\textit{Experiments.} We recall briefly the experimental set-up
described in \cite{Santucci0406}. We use paper which is a
bi-dimensional brittle material with a quasi-linear elastic
stress-strain response until rupture. Samples are fax paper sheets
of size $24 \times 21\,\rm{cm}^2$ manufactured by Alrey. Each sample
has an initial crack at its center and is loaded in a tensile
machine with a constant force $F$ perpendicular to the crack
direction (mode I). The stress intensity factor $K(L) \propto F
\sqrt{L}$, where $L$ is the crack length, determines the stress
magnitude near the crack tip and is the control parameter of crack
growth. For a given initial length $L_i$, sub-critical crack growth
is obtained by choosing $F$ so that $K(L_i)$ is smaller than a
critical threshold $K_c$ corresponding to the material toughness.
During an experiment, $L$ increases, and so does $K(L)$. It will
make the crack accelerate until reaching the critical length $L_c$
for which $K(L_c)=K_c$ and above which a sudden transition to fast
crack propagation occurs. Using a high speed and high resolution
camera (Photron Ultima $1024$), we have determined which part of the
\textit{post mortem} crack interface corresponds to slow or fast
growth, and measure the velocity of the crack in each one. In the
sub-critical regime, the velocity ranges from $10^{-5}$ to
$10^{-2}\,\rm{m.s}^{-1}$. Recording at $4000\,\rm{fps}$, we find a
crack velocity about $300\,\rm{m.s}^{-1}$ in the fast regime. Note
that there are four to seven decades between the two growth regime
velocities.

\begin{figure}[h!]
\includegraphics[width=8.55cm]{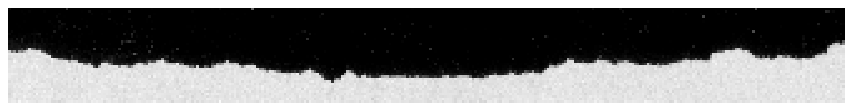}
\includegraphics[width=8.5cm]{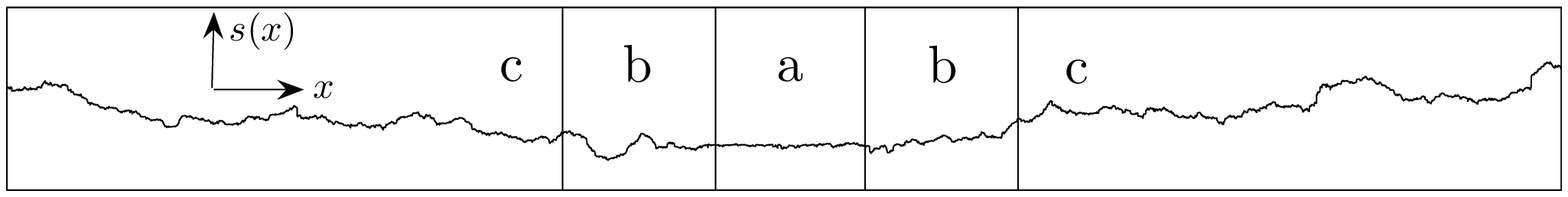}
\caption{Digitized \textit{post mortem} sample and corresponding
extracted front showing three stages: (a) initial crack
($L_i=2\,\rm{cm}$), (b) sub-critical growth, (c) fast growth.}
\label{fig.front}
\end{figure}
\textit{Crack profiles.} \textit{Post mortem} samples are digitized
with a commercial scanner at $1600\,\rm{dpi}$, which corresponds to
a pixel size $a_0=16\,\mu\rm{m}$ close to the typical diameter of
cellulose fibers. Clearly, below this scale, there is no interesting
information about the scaling behaviour of cracks. On
Fig.~\ref{fig.front}, we show an example of a digitized sample
compared with the extracted crack front $s(x)$. We distinguish
between different stages corresponding to: ($a$) the initial crack,
($b$) the sub-critical crack growth and ($c$) the fast crack growth.
We have digitized $51$ fractured samples, obtained with different
forces $(200\,\rm{N}$, $230\,\rm{N}$, $250\,\rm{N}$, $280\,\rm{N}$)
and initial crack sizes ($1\,\rm{cm}$, $2\,\rm{cm}$). Since the
initial crack is centered, each sample give rise to two fronts.
Thus, we have a total of $102$ independent fronts of about $10^3$
points for slow growth and $10^4$ points for fast growth.

\textit{Scale invariance}. Let $\{s(x), x\in \R \}$ be a signal as a
function of a coordinate $x$. Scale invariance of $s$ means that
there is no characteristic scales in the signal. The scaling
properties of $s$ can be characterized by introducing a
multiresolution coefficient $T[s](x,a)$ defined at scale $a$ and
position $x$. Scale invariance implies that the $q$-th order moments
of the multiresolution coefficient follow a power law with exponents
$\zeta(q)$:
 \begin{equation}
M_q\equiv \langle  |T[s](x,a)|^{q}\rangle \sim C |a|^{\zeta(q)}\; ,
\label{structurefunction}
\end{equation}
where the bracket denotes the average over the $x$ space. The
increments over a scale $a$, $T[s](x,a)=s(x+a)-s(x)$, are standard
multiresolution coefficients and the corresponding moments are the
structure functions \cite{Parisi}. It has been shown that a more
general framework for defining multiresolution coefficients is the
wavelet transform \cite{Muzy,Mallat,Simonsen}. When the signal $s$
follows Eq.~(\ref{structurefunction}) with $\zeta(q)$ proportional
to $q$, the signal is monofractal. The complete analysis of the
deviations of $\zeta(q)$ from monofractality can be made through the
multifractal formalism.


\textit{Multifractal analysis}. Multifractal formalism is based on
the mathematical definition of a local singularity exponent $h(x)$.
Using the multiresolution coefficient $T[s](x,a)$, we write at each
position $x$ \cite{Mallat}:
\begin{equation}
T[s](x,a) \sim C |a|^{h(x)}\; ,
\end{equation}
where $h(x)$ is the H\"older exponent or local roughness exponent
describing how singular the signal is at position $x$: the larger
$h(x)$, the smoother $s(x)$. The statistical distribution of the
H\"older exponents is quantified by the singularity spectrum $D(h)$
defined as \cite{Parisi,Mallat}:
\begin{equation}
D(h) =d_H\{ x | h(x)=h\}\;.
\end{equation}
where $d_H$ is the Haussdorf dimension. The probability to observe
an exponent $h$ at scale $a$ is then proportional to $a^{1-D(h)}$.
Thus, the $\zeta(q)$ spectrum can be related to the singularity
spectrum $D(h)$ by a Legendre transform, i.e.
$\zeta(q)=\min_h(1+qh-D(h))$. When $s(x)$ is monofractal, $h(x)$ is
a constant $H$ independent of $x$, $D(H)=1$ and $\zeta(q)= q H$ is
proportional to $q$. Conversely, if $s(x)$ is multifractal, $h(x)$
takes different values at different positions $x$, and $\zeta(q)$ is
not proportional to $q$.

In practice, the values $\zeta(q)$ are obtained by fitting straight
lines (when power law behavior is observed)
 on log-log plots of $M_q$ versus scale $a$ for different moment orders $q$. The
singularity spectrum $D(h)$ is then deducted from $\zeta(q)$. To
verify if $\zeta(q)$ differs from a linear monofractal behavior, one
needs to obtain $\zeta(q)$ for a large range of $q$ values and then
proceed to fit the $\zeta(q)$ curve (for instance, $1 \leq q \leq 8$
in \cite{Bouchbinder06}, see also \cite{Barabasi}).

While this has been the traditional way of estimating $\zeta(q)$, an
alternate method introduced recently in the turbulence literature
\cite{Delour,Chevillard0305}, involves only a few straight line fits
(as low as 3) while still accurately estimating the non-linear
behavior of the $\zeta(q)$ spectrum. To summarize this method, we
start with the general expansion \cite{Delour}:
\begin{eqnarray}
\ln M_q  &=& \sum_{n=1}^{\infty} C_n(a) \frac{q^n}{n!},
\label{eq:derivation_cumulant2}
\end{eqnarray}
where $C_n(a)$ are the cumulants of $Q_a \equiv \ln|T[s](x,a)|$. One
can demonstrate that the first three cumulants are the mean,
standard deviation and skewness of $Q_a$:
\begin{eqnarray}
C_1(a)& = &\langle Q_a\rangle, \nonumber\\
C_2(a)& = & \langle Q_a^2 \rangle - \langle Q_a\rangle^2,\\
C_3(a)& = &\langle Q_a^3 \rangle - 3 \langle Q_a^2\rangle\langle Q_a
\rangle + \langle Q_a \rangle^3.\nonumber
\label{eq:derivation_cumulant4}
\end{eqnarray}
Identifying the first derivative of
Eq.\,(\ref{eq:derivation_cumulant2}) and of the logarithm of
Eq.\,(\ref{structurefunction}) with respect to $\ln(a)$, one finds:
\begin{equation}
\zeta(q)   =  c_1 q + c_2 q^2/2! + c_3 q^3/3!+ \cdots
\label{eq:tauq_cumulant1}
\end{equation}
where $c_i \equiv \rm{d} C_i(a) / \rm{d} \ln(a)$ are constants in
the case of scale invariant signals. It turns out that the average
value of the H\"older exponent is $\langle h \rangle =c_{1}$ and its
variance $\langle h^2 \rangle - \langle h \rangle^2=-c_2/\ln(a)$.
When the multiresolution coefficient $T[s](x,a)$ has Gaussian
statistics (for example, when $s$ is a Brownian motion),
$C_2(a)=\pi^2/8$ and $c_{2}=0$.

The above developments imply that $\zeta(q)$ can be estimated from
linear regressions of the cumulants
 $C_n(a)\,{\rm vs}\,\ln  (a)$  \cite{Delour}. For a monofractal signal,
$c_{n}=0, \forall n\geq 2$ and only one linear regression is needed.
For a multifractal signal, a quadratic $\zeta(q)$ approximation
requires only two linear regressions. In comparison with the
standard method based on the $q$-th order moments, the efficiency of
the cumulant method becomes apparent.

In the following, we will plot only results obtained using the
increments for $T[s](x,a)$. Quantitative comparison with other
methods will be presented afterwards.

\begin{figure}[h!]
\includegraphics[width=8.5cm]{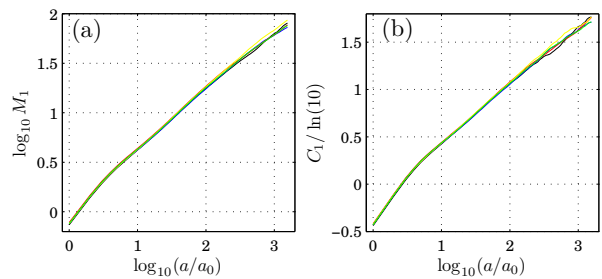}
\caption{(a) $\log_{10} M_1$ and (b) $C_1/\ln(10)$ versus scale
calculated by the increments method for five different couples
$(F,L_i)$ during fast crack growth.} \label{fig.force}
\end{figure}

\textit{Influence of force and initial length.} On
Fig.~\ref{fig.force}, we plot $\log_{10} M_1$ and $C_1/\ln(10)$
versus $\log_{10} (a/a_0)$ for five different couples of values $(F,
L_i)$ during the fast growth stage. We see no dependence of $M_1$
and $C_1$ on the force or the initial crack size. The same
independence is observed for the slow growth stage. These
observations are independent of the definition chosen for
$T[s](x,a)$ which means that our analysis is robust. In order to
improve statistics, we will average the scaling laws obtained for
data with different forces and initial crack lengths. We also notice
that close to the discretization scale $a_0$ the slope becomes close
to unity. This effect can be attributed to the discreteness of the
signal \cite{Mitchell}. In the following, we will concentrate on
larger scales $a> 4 a_0$ ($\simeq$ maximum fiber diameter) where
this effect can be neglected.

\begin{figure}[h!]
\centerline{\includegraphics[width=8.5cm]{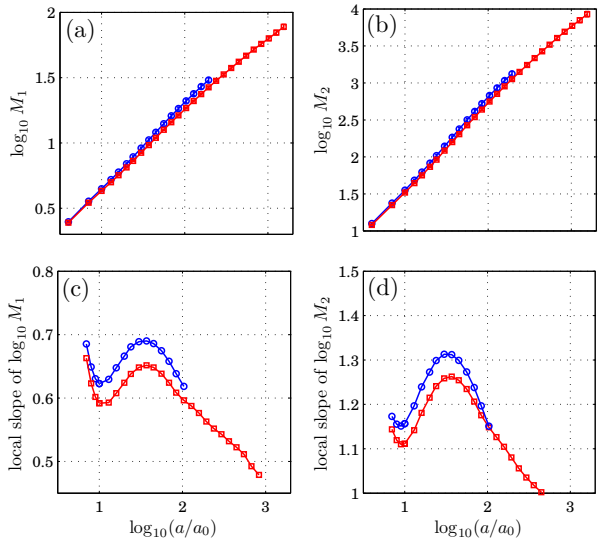}} \caption{(a),
(b): $M_1$ and $M_2$ for slow (circles) and fast (squares) crack
growth. Statistical errors of the mean ($\equiv$ standard deviation
$/\sqrt{N}$ with $N=102$) are smaller than symbol size. (c), (d):
Corresponding local slope with an error comparable to the symbol
size.} \label{fig.lentrap}
\end{figure}

\textit{First and second order moments.} Fig.~\ref{fig.lentrap}(a)
and (b) show log-log plots of $M_1$ and $M_2$ versus $a/a_0$ for
slow and fast growth. The curves are slightly different for slow and
fast crack growth. Another important observation is that they are
not perfectly straight lines. To magnify the difference, we plot the
local slope of those curves (estimated as the mean slope on a
half-decade window) on Fig.~\ref{fig.lentrap}(c) and (d). One can
see an evolution of the slopes with scale, in a similar fashion for
fast or slow cracks. However, we observe a systematic difference
between the slope values which is almost independent of the scale
but depends on the moment order. Also, assuming that the moments are
approximately linear, it is not possible to conclude for a roughness
exponent value since the typical slopes $\zeta(1)$ and $\zeta(2)$
are such that $\zeta(2) \neq 2 \zeta(1)$.

\begin{figure}[h!]
\includegraphics[width=8.5cm]{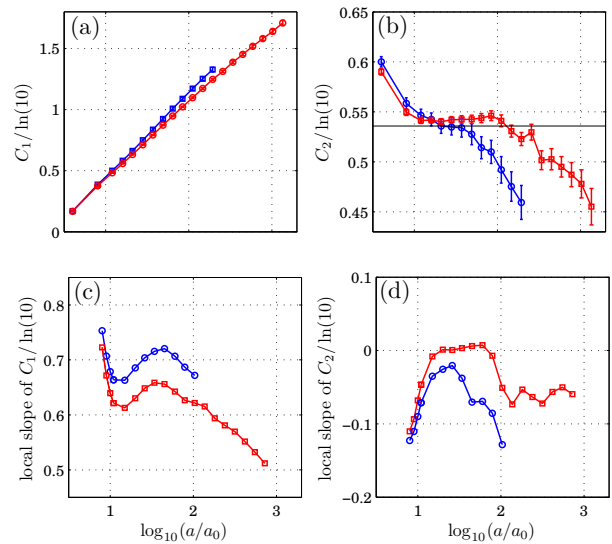}
\caption{(a), (b): $C_1$ and $C_2$ versus scale with statistical
error bars. In (b), the horizontal solid line corresponds to a
signal with Gaussian statistics. (c), (d): Corresponding local slope
of cumulants with an error comparable to the symbol size.}
\label{fig.C2}
\end{figure}

\textit{First and second order cumulants.} On Fig.~\ref{fig.C2}(a)
and (b), we plot $C_1(a)$ and $C_2(a)$ divided by $\rm{ln}(10)$
versus $\rm{log}_{10} (a/a_0)$ for slow and fast crack growth.
$C_1(a)$ is almost linear, but has the same qualitative behavior
than $M_1$ or $M_2$ when looking at the local slope (see
Fig.~\ref{fig.C2}(c)). We will note the local slope  $c_1^F(a)$ for
the fast regime and $c_1^S(a)$ for the slow one. The separation
between fast and slow crack growth is clearer on $C_1$ than on $M_1$
and $M_2$, and the difference in local slopes $\Delta c_1(a)=
c_1^S(a) - c_1^F(a)$ is also more regular. The local extremum
observed for $a/a_0\simeq 10^{1.5}$ ($a\simeq 500\, \mu\rm{m}$) in
Fig.~\ref{fig.C2}(c) suggests the existence of a characteristic
scale for both regimes. In contrast, $C_2(a)$ is hardly a linear
function. We note that there is a range of scales ($10<a/a_0<10^2$)
for which the value of $C_2$ is close to the theoretical value for a
signal with Gaussian statistics (see also \cite{Santucci06}).
However, in this range, the values of the local slope can be
non-zero, especially for the slow regime (see Fig.~\ref{fig.C2}(d)).

\textit{Scaling laws.} From the various plots, we can already
conclude that it is not so easy to find a range of scales for which
clear scaling laws are observed. For $a/a_0\lesssim 10$, the slope
of $C_1(a)$ is changing a lot because we start to feel the
discretization effect previously discussed \cite{Mitchell}. We
clearly see that for $a/a_0\gtrsim 10^2$, the slope of $C_1(a)$ is
again changing significantly and seems to go towards $0.5$. This can
be attributed to a change of statistics at large scales. Thus, if a
scaling law exists, it is observed mainly at intermediate scales
where a small oscillation of the local slope is observed. The same
conclusion can be reached by looking at $C_2(a)$. At large scale,
$C_2(a)$ decreases to values smaller than the Gaussian value that
have no physical meaning. At small scales, $C_2$ is very sensitive
to discretization effects. At intermediate scales ($10<a/a_0<10^2$),
$C_2$ appears quasi-constant. If one tries to estimate the slope
$c_2$, one finds that $c_2$ is very close to zero for the fast crack
growth and between $-0.08$ and $-0.02$ for the slow part.

We notice that the extremum value of $c_2$ ($\sim -0.02$) observed
at intermediate scales for slow crack growth is close to the one
found in \cite{Bouchbinder06} where both regimes were mixed. It is
still difficult to confirm multiscaling since $c_2$ can be
considered constant only over a very small range of scales.
Nevertheless, the non-zero values observed for $c_2$ are important
to consider since they will lead to $\zeta(2)\neq 2 \zeta(1)$.
Indeed, stopping at the second order, Eq.~(\ref{eq:tauq_cumulant1})
predicts: $\zeta(1)=c_1+c_2/2$, and $\zeta(2)=2(c_1+c_2)$.  Thus,
$\zeta(1)$ and $\zeta(2)$ are both influenced by the mean value of
the H\"{o}lder exponent and its variance. In order to give
meaningful information about the scaling properties of the crack, it
is best to estimate directly $c_1$ from $C_1$ since, in the case of
scale invariant signals, $c_1=\langle h \rangle$.

\begin{table}
\caption{\label{Table.Expo} Scaling exponents $c_1 (= \langle h
\rangle$) for fast and slow crack growth, and their difference using
various methods.}
\begin{tabular}{|c|c|c|c|} \hline
  Method  & Fast growth  & Slow growth & Difference \\ \hline
  SF & 0.64 $\pm$ 0.02 & 0.70 $\pm$ 0.02 & 0.06 $\pm$ 0.01\\ \hline
 CWT & 0.65 $\pm$ 0.02 & 0.73 $\pm$ 0.02 & 0.07 $\pm$ 0.01\\ \hline
 WTMM &  0.64 $\pm$ 0.01 & 0.70 $\pm$ 0.02 & 0.06 $\pm$ 0.01  \\ \hline
\end{tabular}
\end{table}

\textit{Scaling exponents.} We have estimated the scaling exponents
in the range of scales for which the scaling laws are reasonably
good. We took $a/a_0 =10$ for the lower cut-off value, and for the
upper cut-off value the point where $C_2$ cross the Gaussian value
($a/a_0=10^2$ for fast cracks, $a/a_0=10^{1.7}$ for slow cracks). In
this scale range, we measure the mean and standard deviation of
$c_1^F$, $c_1^S$ and $\Delta c_1$. We show in Table \ref{Table.Expo}
the mean values $\pm$ the standard deviation using various methods
(SF: structure functions, CWT : Continuous Wavelet Transform
\cite{Parisi} and WTMM : Wavelet Transform Modulus Maxima
\cite{Muzy} using $1^{st}$ derivative of Gaussian). All methods give
a difference between the two growth regimes. For instance, structure
functions give a difference of $0.06 \pm 0.01$ with a roughness
exponent of $0.64 \pm 0.02$ for the fast regime and $0.70 \pm 0.02$
for the slow regime. Note that these values are consistent with the
ones in the literature \cite{Menezes,HorvathSalminen}. The standard
deviation is smaller on the difference than on the scaling exponents
themselves due to the regular shift between local slopes seen on
Fig.~\ref{fig.C2}(c).

\textit{Conclusion.} Whatever are the exact scaling properties of
cracks in paper, we find that the roughness changes when the rupture
mechanism goes from sub-critical to fast crack growth. A decrease of
the roughness exponent due to dynamical instabilities was previously
observed in the case of fast crack growth \cite{Boudet}. It would be
interesting to understand if the smaller roughness in the fast
regime comes from a similar effect. Nevertheless, roughness in the
slow regime is probably controlled more by the material disorder
than by dynamical effects. There seems to be a characteristic scale
$a\simeq 500 \,\mu\rm{m}$ in both growth regimes where the local
slope of moments or cumulants reach an extremum value. Whether this
scale is related to some internal structure of paper, for instance
the fiber length, is an open issue.

\begin{acknowledgments}
We thank S. Ciliberto for fruitful discussions. This work was funded
with grant ANR-05-JCJC-0121-01.
\end{acknowledgments}

\end{document}